\documentclass[conference]{IEEEtran}
\IEEEoverridecommandlockouts
\usepackage{cite}
\usepackage{amsmath,amssymb,amsfonts}
\renewcommand{\theenumi}{\roman{enumi}}
\usepackage{graphicx}
\usepackage{textcomp}
\usepackage{xcolor}
\usepackage{algpseudocode}
\usepackage{amsmath}
\usepackage{amssymb}
\usepackage{algorithm}
\usepackage{algpseudocode}
\usepackage{array}
\usepackage{url}
\usepackage{caption}
\usepackage{subcaption}
\usepackage{cite}
\usepackage{enumerate}
\usepackage{comment}
\usepackage{mdframed}
\usepackage{xcolor}
\usepackage{booktabs}%
\usepackage{soul, color, colortbl}

\usepackage{makecell}
\usepackage{lipsum}
\usepackage{graphicx}
\def\BibTeX{{\rm B\kern-.05em{\sc i\kern-.025em b}\kern-.08em
    T\kern-.1667em\lower.7ex\hbox{E}\kern-.125emX}}
\begin{document}

\title{FA-Net: A Fuzzy Attention-aided Deep Neural Network for Pneumonia Detection in Chest X-Rays}

\author{\IEEEauthorblockN{Ayush Roy}
\IEEEauthorblockA{\textit{Dept. of Electrical Engineering} \\
\textit{Jadavpur University}\\
Kolkata, India \\
aroy80321@gmail.com}
\and
\IEEEauthorblockN{Anurag Bhattacharjee}
\IEEEauthorblockA{\textit{Dept. of Computer Science and Engineering} \\
\textit{Kalinga Institute of Industrial Technology}\\
Bhubaneswar, India \\
anuragdgp@gmail.com}
\and
%\IEEEauthorblockN{Diego Oliva}
%\IEEEauthorblockA{\textit{Dept. of Computer Science} \\
%\textit{Universidad de Guadalajara}\\
%Jalisco, Mexico \\
%diego.oliva@cucei.udg.mx}
%\and
%\and
\IEEEauthorblockN{Diego Oliva}
\IEEEauthorblockA{%\textit{} \\
\textit{Universidad de Guadalajara, CUCEI}\\
Guadalajara, Mexico \\
diego.oliva@cucei.udg.mx}
\and
\IEEEauthorblockN{Oscar Ramos-Soto}
\IEEEauthorblockA{%\textit{} \\
\textit{Universidad de Guadalajara, CUCEI}\\
Guadalajara, Mexico \\
oscar.ramos9279@alumnos.udg.mx\\}
\and
\IEEEauthorblockN{Francisco J. Alvarez-Padilla}
\IEEEauthorblockA{%\textit{} \\
\textit{Universidad de Guadalajara, CUCEI}\\
Guadalajara, Mexico \\
francisco.alvarez@academicos.udg.mx}
\and
\IEEEauthorblockN{Ram Sarkar}
\IEEEauthorblockA{\textit{Dept. of Computer Science and Engineering} \\
\textit{Jadavpur University}\\
Kolkata, India \\
ramjucse@gmail.com}

}

\maketitle

\begin{abstract}
Pneumonia is a respiratory infection caused by bacteria, fungi, or viruses. It affects many people, particularly those in developing or underdeveloped nations with high pollution levels, unhygienic living conditions, overcrowding, and insufficient medical infrastructure. Pneumonia can cause pleural effusion, where fluids fill the lungs, leading to respiratory difficulty. Early diagnosis is crucial to ensure effective treatment and increase survival rates. Chest X-ray imaging is the most commonly used method for diagnosing pneumonia. However, visual examination of chest X-rays can be difficult and subjective. In this study, we have developed a computer-aided diagnosis system for automatic pneumonia detection using chest X-ray images. We have used DenseNet-121 and ResNet50 as the backbone for the binary class (pneumonia and normal) and multi-class (bacterial pneumonia, viral pneumonia, and normal) classification tasks, respectively. We have also implemented a channel-specific spatial attention mechanism, called Fuzzy Channel Selective Spatial Attention Module (FCSSAM), to highlight the specific spatial regions of relevant channels while removing the irrelevant channels of the extracted features by the backbone. We evaluated the proposed approach on a publicly available chest X-ray dataset, using binary and multi-class classification setups. Our proposed method achieves accuracy rates of 97.15\% and 79.79\% for the binary and multi-class classification setups, respectively. The results of our proposed method are superior to state-of-the-art (SOTA) methods. The code of the proposed model will be available at: \url{https://github.com/AyushRoy2001/FA-Net}
\end{abstract}


\section{Introduction}
\begin{figure}[ht]
    \centering
    \includegraphics[width=0.9\linewidth]{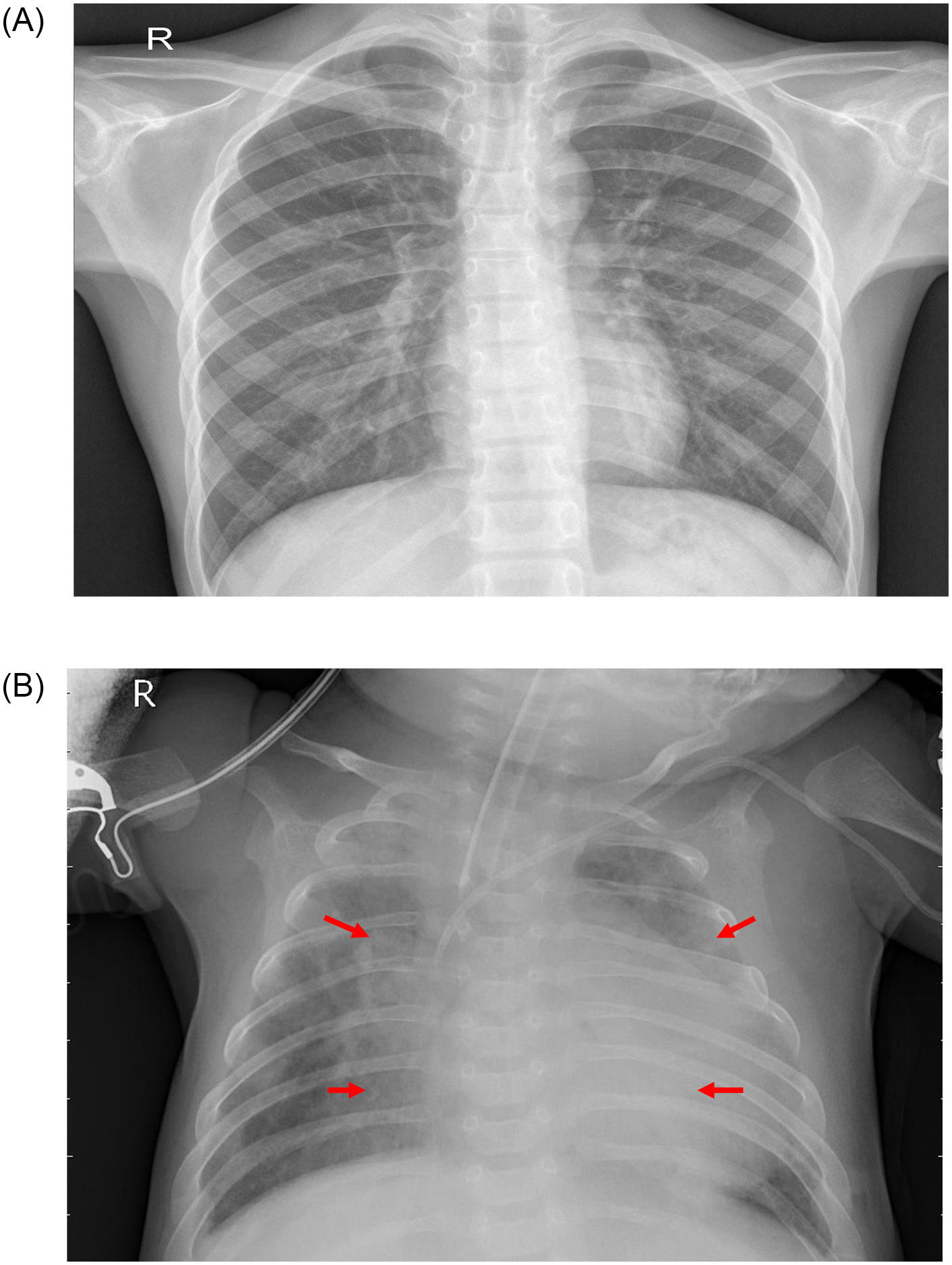}
    \caption{An example of two X-ray plates that display a healthy lung (A) and a pneumonic lung (B). In (B), the red arrows indicate white infiltrates which is a distinguishing feature of pneumonia.} 
    \label{fig:intro}
\end{figure}

Pneumonia, an acute pulmonary infection, is a leading cause of mortality, especially among children under five years old, accounting for over 15\% of deaths globally \cite{who2019pneumonia}. It can result from bacterial, viral, or fungal infections, leading to inflammation of the lung's air sacs and pleural effusion, where fluid accumulates in the lung. The prevalence of pneumonia is higher in underdeveloped and developing nations due to factors like overpopulation, pollution, and inadequate medical resources. Early diagnosis and management are crucial in preventing fatalities. Radiological imaging, such as X-rays, computed tomography (CT) scans, or magnetic resonance imaging (MRI), is commonly used to diagnose pneumonia. Chest X-rays reveal characteristic white spots, called infiltrates, indicative of pneumonia. However, the interpretation of these X-rays can vary subjectively \cite{neuman2012variability,williams2013variability}, leading to the need for automated detection systems. This can be seen in Fig. \ref{fig:intro}.

Deep learning models are advanced solutions that use multiple layers of artificial neural networks for various purposes and have proven their edge over machine learning algorithms in various tasks \cite{goyal2023detection, khan2021intelligent, yi2023identification}. They have shown great speech recognition, language translation, and image processing results. Researchers have utilized deep learning-based models to automate pneumonia diagnosis using chest X-rays or CT scans. Mabrouk et al. \cite{mabrouk2022pneumonia} developed a feature stacking technique to combine three well-known models, DenseNet169, MobileNetV2, and Vision Transformer, for pneumonia classification. To address issues related to generalization, dataset size, and time complexity, Vrbancic, and Podgorelec \cite{vrbanvcivc2022efficient} created an ensemble method based on stochastic gradient descent (SGD) with warm restarts (SGDRE). This method exploits the generalization capabilities of ensemble methods and SGD with a warm restarts mechanism to obtain a diverse group of classifiers in a single training process, spending the same or less training time than a single CNN model. Mabrouk et al. \cite{mabrouk2023ensemble} developed an Ensemble method-based Federated Learning (EFL) framework for pneumonia identification using Chest X-ray images. This framework allows researchers to gather knowledge about personal medical data without publishing this data, maintaining privacy while working in a decentralized and collaborative manner. Recently, Ali et al. \cite{ali2024pneumonia} showed that EfficientNet-V2L is an effective standard convolutional neural network (CNN) architecture, which can handle the image quality issue of X-rays. RAPID-Net \cite{dabre2024rapid} is another method that addresses various issues related to pneumonia diagnosis. It uses thin stacking of convolutional layers with faster convergence and suitable preprocessing and augmentation techniques. This method addresses issues like dataset imbalance, limited quality of chest radiograph, pathological deformities, X-ray imaging inhomogeneities, gross background noise, overlapped patterns of opacities, and anatomical alterations caused by misaligned body positioning. It extracts significant local and global statistical features for pneumonia cloud detection. 

However, existing methods have some issues that need to be addressed. One of the main problems is that they do not filter out irrelevant features, which the deep learning model learns. This negatively affects the performance of the models. Additionally, there is a need for a spatial attention mechanism, which can highlight the spatial regions responsible for correct classification. Considering the above facts, we have proposed a computer-aided diagnosis (CAD) system utilizing deep transfer learning models for pneumonia detection from chest X-ray images. The dataset presented by Kermany et al. \cite{kermany2018large} is a chest X-ray image set for binary classification with ``normal'' and ``pneumonia'' classes. The ``pneumonia'' one can be further divided into two more subclasses, i.e., ``bacterial pneumonia'' and ``viral pneumonia''. We have tested our model on both these setups to demonstrate the robustness of the proposed model and capture intricate differences among the classes of pneumonia. The highlighting points of this work include:

\begin{enumerate}
    \item It considers DenseNet-121 and ResNet50 as the backbone for binary (2-class) and multi-class (3-class) classification tasks, respectively. 
    \item The extracted features from the backbone are then enriched using a spatial and channel attention module of the Fuzzy Channel Selective Spatial Attention Module (FCSSAM) before filtering out the relevant channels of the enriched feature map.
    \item This model enables accurate classification of chest X-ray images and showcases an increase in accuracy (+0.52\%), AUC (+2.74\%), and recall (+1.1\%) in comparison to the SOTA methods for the 2-class setup. For the multi-class setup, the proposed model demonstrates superior performance compared to the standard Convolutional Neural Networks (CNNs) in terms of accuracy (+3.51\%), precision (+2.35\%), recall (+3.51\%), AUC (+3.03\%), and F1-score (+2.89\%).
\end{enumerate}

\section{Methodology}
\begin{figure*}[ht]
    \centering
    \includegraphics[width=\linewidth]{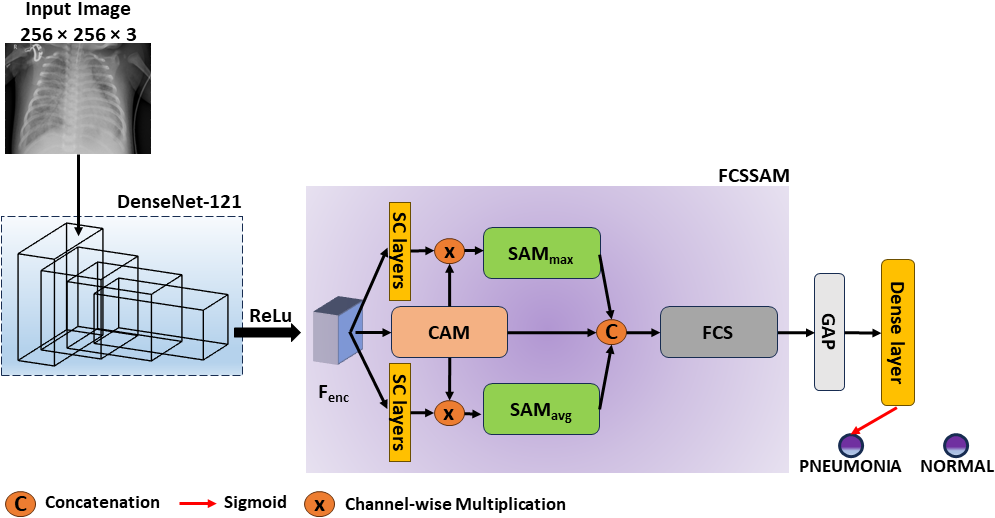}
    \caption{A pictorial representation of the proposed pneumonia detection model using chest X-ray images.} 
    \label{fig:overall}
\end{figure*}

The proposed model is illustrated in the block diagram presented in Fig. \ref{fig:overall}. To extract the feature $F_{enc}$ of dimensions $H \times W \times C$, we use the best performing CNN, i.e., DenseNet169 as the backbone for binary classification and ResNet50 as the backbone for multi-class classification. The input image has dimensions $256 \times 256 \times 3$. We then pass this feature vector on to three branches that are treated by Channel Attention Module (CAM) and Spatial Attention Modules ($SAM_{avg}$ and $SAM_{max}$). These modules highlight the specific regions and channels that enhance the model's performance. The SC layers are two consecutive Separable Convolution layers of kernel size 1 and 3, and $C$ number of filters, respectively. The output features from $SAM_{avg}$ and $SAM_{max}$ are concatenated to produce a feature map of dimensions $H \times W \times 2C$. This feature map serves as an input for the Fuzzy Channel Selection (FCS) module to generate a feature map with the top $C$ channels according to the assigned learnable weights. Finally, the output from the FCS module is flattened using the Global Average Pooling (GAP) layer. The classification layer then treats this flattened feature to predict the output of the image.

\subsection{Fuzzy Channel Selective Spatial Attention Module}
The FCSSAM consists of three components, as shown in Fig. \ref{fig:overall}, that are described below:

\textbf{Channel Attention Module (CAM):} The CAM assigns weights to the channels of feature maps, which enhance particular channels that contribute more towards improving model performance. The output feature map of the 1D channel attention network is denoted as $F_{CAM}$ and has a dimension of $C \times 1 \times 1$. $F_{CAM}$ is a combination of two features, $F_{exavg}$ and $F_{exmax}$, which can be defined using Eq. \ref{eq:cam_avg} and Eq. \ref{eq:cam_max}, respectively:

\begin{equation}
    F_{exavg} = D_{2}(ReLU(D1(GAP(F_{enc})))
    \label{eq:cam_avg}
\end{equation}

\begin{equation}
    F_{exmax} = D_{2}(ReLU(D1(GMP(F_{enc})))
    \label{eq:cam_max}
\end{equation}

In the above equations, $D_1$ and $D_2$ are dense layers with units $C/r$ and $C$, respectively. GAP and GMP are the Global Average Pooling and Global Max Pooling layers. The ratio $r$ denotes the amount the features are squeezed to $C/r$ from $C$ and then excited again to its original dimension $C$. This is why it is also known as squeeze and excitation attention. $F_{CAM}$ can be obtained using $F_{exavg}$ and $F_{exmax}$, as shown in Eq. \ref{eq:cam3}, where $\sigma$ denotes the sigmoid activation, and `$+$' denotes an element-wise addition:

\begin{equation}
    F_{CAM} = \sigma(F_{exmax} + F_{exavg})
    \label{eq:cam3}
\end{equation}

\begin{figure}[ht]
    \centering
    \includegraphics[width=\linewidth]{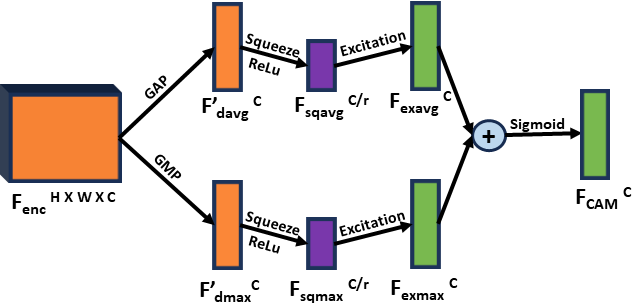}
    \caption{The Channel Attention Module (CAM).} 
    \label{fig:CAM}
\end{figure}

\textbf{Spatial Attention Module (SAM):} The SAM enhances important spatial regions in an input feature map F by selectively amplifying informative areas while suppressing less relevant ones. The $SAM_{avg}$ and $SAM_{max}$ utilizes two different feature maps, $F_{avg}$ and $F_{max}$ respectively, as inputs. The first step involves computing $F_{max}$ and $F_{avg}$ of dimensions $H \times W \times 1$ by performing max and average pooling respectively on the spatial dimensions of the input feature map $F$ of dimensions $H \times W \times C$. Subsequently, a convolutional layer with a kernel size of $7 \times 7$ (denoted as $f^{7 \times 7}$) and sigmoid activation is applied to $F_{max}$ and $F_{avg}$ to obtain attention weights $F_{SAMmax}$ and $F_{SAMavg}$ of dimension $H \times W \times 1$ as shown in Eq. \ref{eq:sam_max} and Eq. \ref{eq:sam_avg}, respectively. The sigmoid activation, denoted by $\sigma$, constrains the range output [0, 1].

\begin{equation}
    F_{SAMmax} = \sigma(f^{7 \times 7}(F_{max}))
    \label{eq:sam_max}
\end{equation}

\begin{equation}
    F_{SAMavg} = \sigma(f^{7 \times 7}(F_{avg}))
    \label{eq:sam_avg}
\end{equation}

Finally, each element of $F$ is multiplied by the corresponding attention weight from $F_{SAMmax}/F_{SAMavg}$ element-wise highlighting important spatial regions in the original feature map based on the learned attention weights. This enriches the representation of the input feature map $F$ by adaptively assigning importance to different spatial regions. The output features from $SAM_{avg}$ and $SAM_{max}$ are concatenated together to produce $F_{concat}$ of dimensions $H \times W \times 2C$

\begin{figure}[ht]
    \centering
    \includegraphics[width=0.9\linewidth]{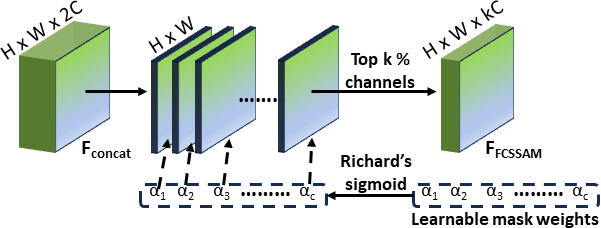}
    \caption{The Fuzzy Channel Selection (FCS) module.} 
    \label{fig:FCS}
\end{figure}

\textbf{Fuzzy Channel Selection (FCS):} After obtaining $F_{concat}$ from SAM, relevant channels are selected by FCS. This selection process ensures that only the most useful channels or combinations of channels are retained, while redundant and irrelevant ones are pruned. This reduces the computational burden by decreasing the feature dimensions from $H \times W \times 2C$ to $H \times W \times kC$, as illustrated in Fig. \ref{fig:FCS}. 

\begin{figure}[ht]
    \centering
    \includegraphics[width=\linewidth]{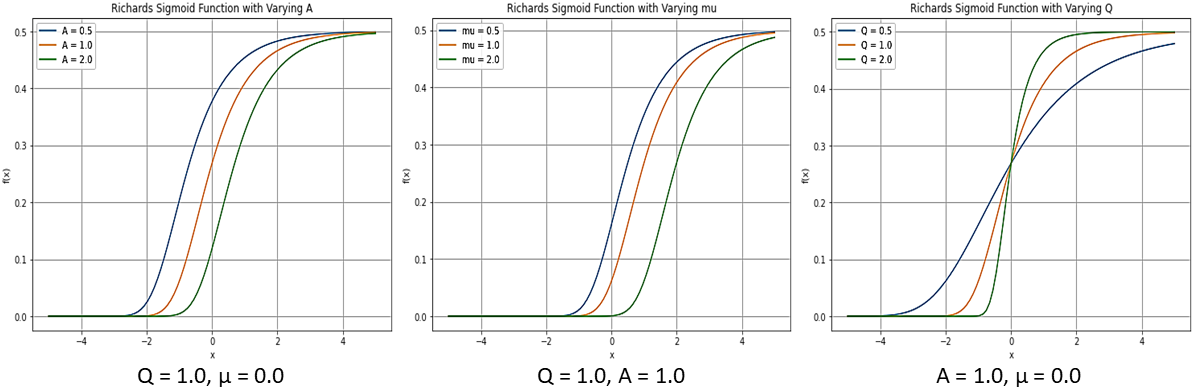}
    \caption{The illustration of the effect in the curve of Richards sigmoid function with adjustments in values of $A$, $Q$, and $\mu$.} 
    \label{fig:RS}
\end{figure}

A learnable mask is assigned to each channel of $F_{concat}$, consisting of weights $\alpha_{1}$, $\alpha_{2}$, $\alpha_{3}$, and so on up to $\alpha_{C}$. These weights are controlled using the Richards sigmoid function to determine the importance of each channel. Based on the values of $\alpha_{i}$, the channels are sorted, and the top $k$\% of the channels are retained while the others are removed. This ensures that only the most relevant channels are selected. The equation of the Richards sigmoid function is defined by Eq. \ref{eq:sigmoid}, where $A$ is a scaling parameter, $Q$ controls the steepness, and $\mu$ is the location parameter. $A$, $Q$, and $\mu$ are made learnable to adjust the sigmoid curve following the mask weights. An effect of the change in curve concerning the changes in values of $A$, $Q$, and $\mu$ is shown in Fig. \ref{fig:RS}. With an increased linearity of the curve, the channels get uniform attention scores in increasing order, whereas a steeper curve ensures higher values for important channels while assigning lower values to the irrelevant ones. 

\begin{equation}
f(x) = \frac{1}{1 + e^{A \cdot e^{-Q \cdot (\alpha - \mu)}}}
\label{eq:sigmoid}
\end{equation}

\section{Results}
\subsection{Experimental Details}
\begin{figure}[ht]
    \centering
    \includegraphics[width=\linewidth]{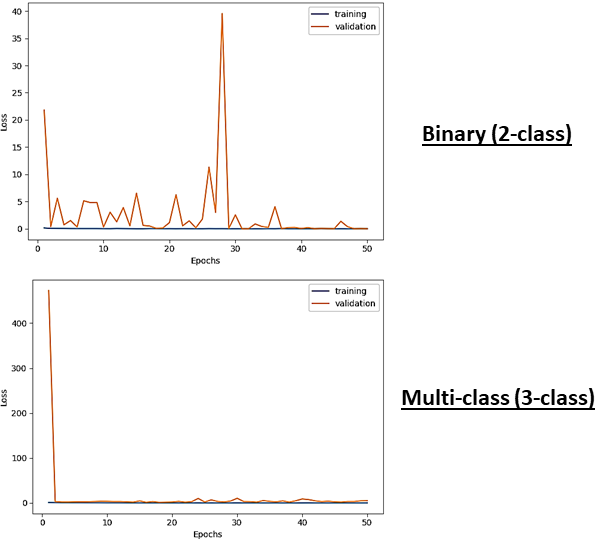}
    \caption{Training and validation loss curves of the binary and multi-class setups.} 
    \label{fig:loss}
\end{figure}

We have researched using the Kermany dataset \cite{kermany2018large} to train and evaluate our model and all the models used for experimentation. The dataset comprises 5,856 chest X-ray images from adults and children, with an uneven distribution between the ``Pneumonia'' and ``Normal'' classes. This class imbalance presented challenges in achieving SOTA results. The testing set has 624 images, with 234 labeled as ``Normal" and 390 labeled as ``Pneumonia". The training set has 5,232 images, with 1,349 labeled as ``Normal" and 3,883 labeled as ``Pneumonia". Data augmentation techniques, including random rotations, shifts, flips, and zooms to the original images, have been applied to generate additional training samples to improve the model's robustness and prevent overfitting. We have used 10\% of the training set images for validation. Since the pneumonia images have two sub-classes, ``Bacterial Pneumonia" and ``Viral Pneumonia," we have developed a three-class setup to produce results. All experiments have been conducted with an image size of $256 \times 256 \times 3$, a learning rate of 0.0001, the Adam optimizer, and a batch size of 48 for all experiments. The cross-entropy loss function is applied to train the model over 50 epochs in an NVIDIA TESLA P100 GPU. Also, we have evaluated it using standard metrics in the TensorFlow framework, including accuracy (Acc), precision (Pre), recall (Rec), and F1-score (F1). Training and validation loss curves of the proposed model for 50 epochs are shown in Fig. \ref{fig:loss} for multi-class and binary setup.

\subsection{Ablation Studies}
\begin{figure}[ht]
    \centering
    \includegraphics[width=\linewidth]{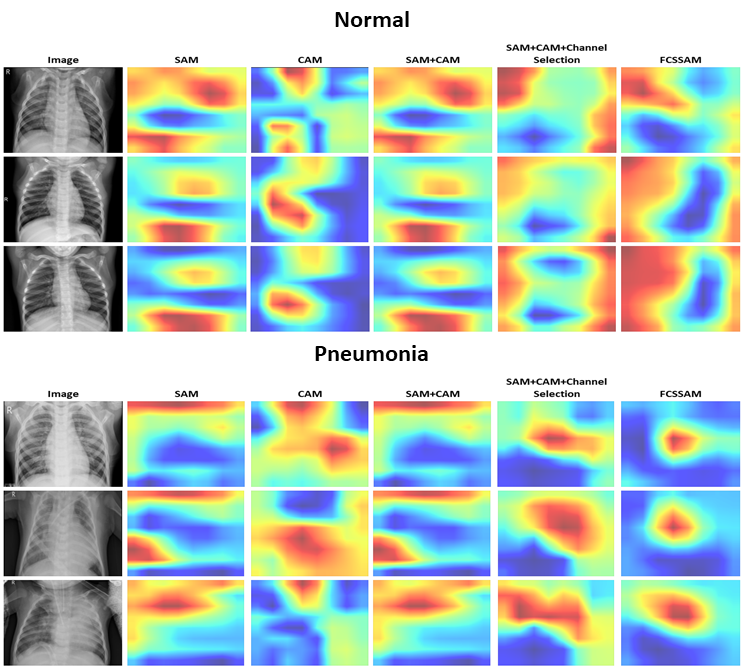}
    \caption{Heatmap of the various versions of FCSSAM (as shown in Table. \ref{ablation_multiclass}) used for ``Normal" and ``Pneumonia" classes.} 
    \label{fig:heat_ablation}
\end{figure}

We have conducted a series of experiments to figure out the best-performing CNN for multi-class and binary classification tasks. All the scores in all tables are expressed as \% and the highest values (indicating highest performance) are marked in bold. It is evident from Table \ref{ablation_binary} and Table \ref{ablation_multiclass}, which show transfer learning result for different CNN models performance for 2-class and 3-class classification, that DenseNet169 and ResNet50 are the best-performing CNN models, respectively. Keeping DenseNet169 as the backbone, we have tested the various configurations of the FCSSAM to demonstrate the importance of each component. As noticeable in Table \ref{ablation_FCSSAM}, the combination of CAM and SAM boosts the recall significantly while the precision deteriorates. Introducing the FCS module removes the redundant channels while keeping the most relevant ones to improve the recall further. The learnable Richards sigmoid in the FCSSSAM module provides the necessary weight adjustment to boost precision and recall, thus producing a higher F1 score. Experimentally, the value of $k$ for FCSSAM is found to be 80\%. The performance enhancement after the application of FCSSAM on the best performing CNN, i.e., ResNet50, for the multi-class setup is shown in Table \ref{ablation_multiclass}, where the accuracy and F1-score increase by 3.51\% and 2.89\%, respectively. Heatmaps of each component of FCSSAM for each of the classes are shown in Fig. \ref{fig:heat_ablation}, where the differences in the focus of the components of FCSSAM can be seen.

\begin{table}[hbt]
    \centering
    \caption{Transfer learning results of CNN models for 2-class classification.}
    \begin{tabular}{@{\extracolsep\fill}cccccc@{}}       
        \hline
        \textbf{Model} & \textbf{Acc} & \textbf{Pre} & \textbf{Rec} & \textbf{AUC} & \textbf{F1}\\
        \hline
        DenseNet169 & \textbf{92.95} & \textbf{91.39} & 97.95 & \textbf{96.41} & \textbf{94.55} \\
        MobileNet-v2 & 84.78 & 86.97 & 88.97 & 89.41 & 87.96 \\
        Xception & 75.80 & 72.09 & \textbf{100.00} & 73.68 & 83.78 \\
        ResNet-50 & 68.59 & 66.55 & \textbf{100.00} & 77.35 & 79.92 \\
        VGG-16 & 75.80 & 72.50 & 98.72 & 85.35 & 83.60 \\
        NasNet-Mobile & 79.81 & 75.98 & 98.97 & 75.09 & 85.97 \\
        Inception-ResNet-v2 &	81.89 & 77.64 & 99.74 & 94.82 & 87.32 \\
        \hline
    \label{ablation_binary}
    \end{tabular}
\end{table}

\begin{table}[hbt]
    \centering
    \caption{Transfer learning results of CNN models compared with the proposed method for 3-class classification.}
    \begin{tabular}{@{\extracolsep\fill}cccccc@{}}       
        \hline
        \textbf{Model} & \textbf{Acc} & \textbf{Pre} & \textbf{Rec} & \textbf{AUC} & \textbf{F1}\\
        \hline
        DenseNet169 & 74.36 & 74.48 & 74.36 & 87.18 & 74.53 \\
        MobileNet-v2 & 58.81 & 58.81 & 58.81 & 71.00 & 57.40 \\
        Xception & 74.84 & 74.76 & 74.52 & 86.65 & 74.69 \\
        ResNet-50 & 76.28 & 76.61 & 76.12 & 87.91 & 76.51 \\
        VGG-16 & 71.47 & 74.39 & 68.43 & 82.92 & 70.64 \\
        NasNet-Mobile & 66.51 & 66.61 & 66.51 & 78.33 & 66.42 \\
        Inception-ResNet-v2 & 74.04 & 74.04 & 74.04 & 85.40 & 74.22 \\
        \textbf{Proposed} & \textbf{79.79} & \textbf{78.96} & \textbf{79.63} & \textbf{90.94} & \textbf{79.40} \\
        \hline
    \label{ablation_multiclass}
    \end{tabular}
\end{table}

\begin{table}[hbt]
    \centering
    \caption{Performance of various components of FCSSAM for 2-class classification.}
    \begin{tabular}{@{\extracolsep\fill}cccccc@{}}       
        \hline
        \textbf{Model} & \textbf{Acc} & \textbf{Pre} & \textbf{Rec} & \textbf{AUC} & \textbf{F1}\\
        \hline
        DenseNet169+CAM & 93.86 & 91.04 & 95.36 & 95.26 & 94.09 \\
        DenseNet169+SAM & 92.60 & 92.85 & 93.74 & 95.71 & 93.60 \\
        DenseNet169+SAM+CAM & 94.60 & 91.85 & 96.74 & 95.71 & 94.20 \\
        DenseNet169+SAM with FCS & 94.34 & 90.79 & 97.38 & 95.27 & 94.56 \\
        DenseNet169+CSSAM & 96.87 & 93.29 & 98.13 & 97.85 & 96.99 \\
        DenseNet169+FCSSAM & \textbf{97.15} & \textbf{94.28} & \textbf{98.67} & \textbf{97.89} & \textbf{97.30} \\
        \hline
    \label{ablation_FCSSAM}
    \end{tabular}
\end{table}

\subsection{SOTA Comparison}
\begin{figure}[ht]
    \centering
    \includegraphics[width=\linewidth]{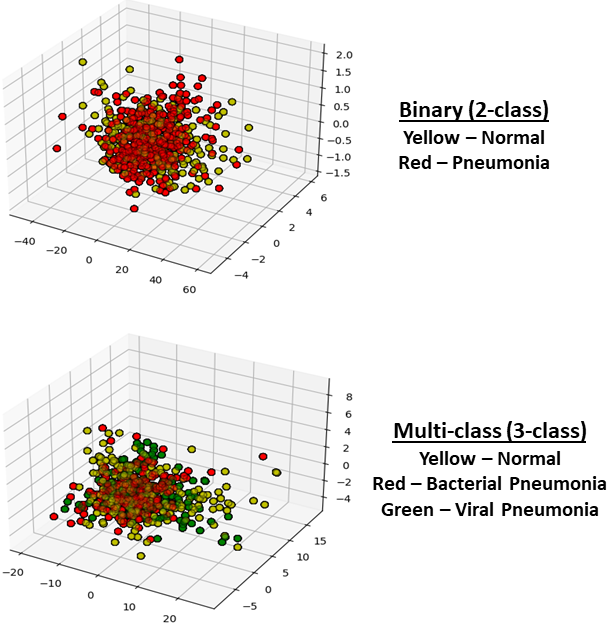}
    \caption{An illustration of the feature representation of the GAP layer of the proposed model for binary (2-class) and multi-class (3-class) setups.} 
    \label{fig:space}
\end{figure}

\begin{figure}[ht]
    \centering
    \includegraphics[width=\linewidth]{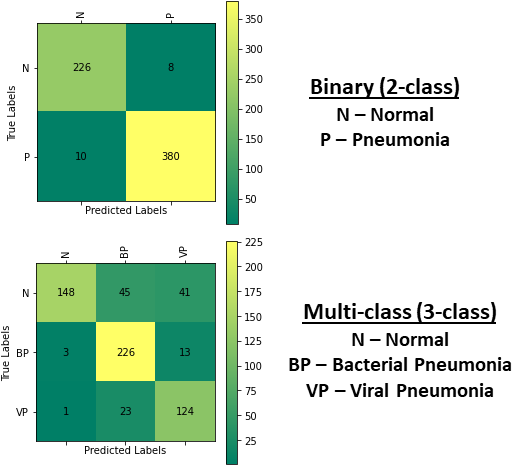}
    \caption{Confusion matrices of the proposed model for binary (2-class) and multi-class (3-class) setups.} 
    \label{fig:cm}
\end{figure}

Our proposed model outperforms existing methods in terms of accuracy and F1-score, as shown in Table \ref{sota}. The proposed model showcases an increase in accuracy (+0.52\%), AUC (+2.74\%), and recall (+1.1\%) in comparison to the SOTA methods. Fig. \ref{fig:space} demonstrates the feature representation of the proposed model after the GAP layer. The flattened features are converted to a 3-dimensional vector using principal component analysis (PCA) and then plotted. The plot demonstrates the features of different classes lie distinctly in the latent space for an efficient classification. The confusion matrices of the proposed model are shown in Fig. \ref{fig:cm} for both the setups, i.e., multiclass and binary.

\begin{table}[hbt]
    \centering
    \caption{Comparison with the SOTA models for 2-class classification.}
    \begin{tabular}{@{\extracolsep\fill}cccccc@{}}       
        \hline
        \textbf{Model} & \textbf{Acc} & \textbf{Pre} & \textbf{Rec} & \textbf{AUC} & \textbf{F1}\\
        \hline
        Mabrouk et al.~\cite{mabrouk2022pneumonia} & 93.91 & 93.96 & 92.99 & - & 93.43 \\
        Vrbancic and Podgorelec~\cite{vrbanvcivc2022efficient} & 96.26 & 97.32 & 97.57 & 95.15 & \textbf{97.44}  \\
        Mabrouk et al.~\cite{mabrouk2023ensemble} & 96.63 & \textbf{97.86} & 96.63 & - & 95.62 \\
        EfficientNet-V2L~\cite{ali2024pneumonia} & 94.02 & 94.40 & 97.24 & - & 95.80 \\
        RAPID-Net~\cite{dabre2024rapid} & 94.60 & 93.18 & 94.62 & - & 93.89 \\
        \textbf{Proposed} & \textbf{97.15} & 94.28 & \textbf{98.67} & \textbf{97.89} & 97.30 \\
        \hline
    \label{sota}
    \end{tabular}
\end{table}

\section{Conclusion}
Pneumonia, an actual public health issue, presents a substantial health struggle around the world. This condition can lead to lung fluid accumulation and respiratory complications. Early detection is a key challenge for successful treatment and increased survival prospects. While chest X-ray scans are commonly employed to diagnose pneumonia, their interpretation can be difficult.

In this work, we have proposed a CAD system for automatic pneumonia detection using chest X-ray images. We have used DenseNet-121 and ResNet50 to perform binary class and multi-class classification tasks, respectively. Additionally, we have implemented a channel-specific spatial attention mechanism, called FCSSAM, to highlight the specific spatial regions of relevant channels while removing the irrelevant channels of the extracted features by the backbone. We have evaluated the proposed approach on a publicly available chest X-ray dataset provided by Kermany et al. \cite{kermany2018large}, using both binary class and multi-class classification setups. Our proposed method outperforms SOTA methods. However, it has scope for improvement considering a lower F1-score and precision concerning the SOTA models. Major reasons for these errors are the class imbalance and poor data quality that need to be addressed in the future. A more robust feature pruning method is essential to boost the performance further. Also, consideration of global semantic features and the local ones is a potential area to research further. 

\section{Acknowledgement}
We are thankful to the Center for Microprocessor Applications for Training Education and Research (CMATER) research laboratory of the Computer Science and Engineering Department, Jadavpur University, Kolkata, India for providing infrastructural support.

\bibliographystyle{IEEEtran}
\bibliography{IEEEfull.bib}

\begin{thebibliography}{10}
\providecommand{\url}[1]{#1}
\csname url@samestyle\endcsname
\providecommand{\newblock}{\relax}
\providecommand{\bibinfo}[2]{#2}
\providecommand{\BIBentrySTDinterwordspacing}{\spaceskip=0pt\relax}
\providecommand{\BIBentryALTinterwordstretchfactor}{4}
\providecommand{\BIBentryALTinterwordspacing}{\spaceskip=\fontdimen2\font plus
\BIBentryALTinterwordstretchfactor\fontdimen3\font minus \fontdimen4\font\relax}
\providecommand{\BIBforeignlanguage}[2]{{%
\expandafter\ifx\csname l@#1\endcsname\relax
\typeout{** WARNING: IEEEtran.bst: No hyphenation pattern has been}%
\typeout{** loaded for the language `#1'. Using the pattern for}%
\typeout{** the default language instead.}%
\else
\language=\csname l@#1\endcsname
\fi
#2}}
\providecommand{\BIBdecl}{\relax}
\BIBdecl

\bibitem{who2019pneumonia}
{World Health Organization}, ``Pneumonia,'' \url{https://www.who.int/news-room/fact-sheets/detail/pneumonia}, 2019.

\bibitem{neuman2012variability}
M.~I. Neuman, E.~Y. Lee, S.~Bixby, S.~Diperna, J.~Hellinger, R.~Markowitz, S.~Servaes, M.~C. Monuteaux, and S.~S. Shah, ``Variability in the interpretation of chest radiographs for the diagnosis of pneumonia in children,'' \emph{Journal of hospital medicine}, vol.~7, no.~4, pp. 294--298, 2012.

\bibitem{williams2013variability}
G.~J. Williams, P.~Macaskill, M.~Kerr, D.~A. Fitzgerald, D.~Isaacs, M.~Codarini, M.~McCaskill, K.~Prelog, and J.~C. Craig, ``Variability and accuracy in interpretation of consolidation on chest radiography for diagnosing pneumonia in children under 5 years of age,'' \emph{Pediatric Pulmonology}, vol.~48, no.~12, pp. 1195--1200, 2013.

\bibitem{goyal2023detection}
S.~Goyal and R.~Singh, ``Detection and classification of lung diseases for pneumonia and covid-19 using machine and deep learning techniques,'' \emph{Journal of Ambient Intelligence and Humanized Computing}, vol.~14, no.~4, pp. 3239--3259, 2023.

\bibitem{khan2021intelligent}
W.~Khan, N.~Zaki, and L.~Ali, ``Intelligent pneumonia identification from chest x-rays: A systematic literature review,'' \emph{IEEE Access}, vol.~9, pp. 51\,747--51\,771, 2021.

\bibitem{yi2023identification}
R.~Yi, L.~Tang, Y.~Tian, J.~Liu, and Z.~Wu, ``Identification and classification of pneumonia disease using a deep learning-based intelligent computational framework,'' \emph{Neural Computing and Applications}, vol.~35, no.~20, pp. 14\,473--14\,486, 2023.

\bibitem{mabrouk2022pneumonia}
A.~Mabrouk, R.~P. Diaz~Redondo, A.~Dahou, M.~Abd~Elaziz, and M.~Kayed, ``Pneumonia detection on chest x-ray images using ensemble of deep convolutional neural networks,'' \emph{Applied Sciences}, vol.~12, no.~13, p. 6448, 2022.

\bibitem{vrbanvcivc2022efficient}
G.~Vrban{\v{c}}i{\v{c}} and V.~Podgorelec, ``Efficient ensemble for image-based identification of pneumonia utilizing deep cnn and sgd with warm restarts,'' \emph{Expert Systems with Applications}, vol. 187, p. 115834, 2022.

\bibitem{mabrouk2023ensemble}
A.~Mabrouk, R.~P.~D. Redondo, M.~Abd~Elaziz, and M.~Kayed, ``Ensemble federated learning: An approach for collaborative pneumonia diagnosis,'' \emph{Applied Soft Computing}, vol. 144, p. 110500, 2023.

\bibitem{ali2024pneumonia}
M.~Ali, M.~Shahroz, U.~Akram, M.~F. Mushtaq, S.~C. Altamiranda, S.~A. Obregon, I.~D. L.~T. D{\'\i}ez, and I.~Ashraf, ``Pneumonia detection using chest radiographs with novel efficientnetv2l model,'' \emph{IEEE Access}, 2024.

\bibitem{dabre2024rapid}
K.~Dabre, S.~L. Varma, and P.~B. Patil, ``Rapid-net: Reduced architecture for pneumonia in infants detection using deep convolutional framework using chest radiograph,'' \emph{Biomedical Signal Processing and Control}, vol.~87, p. 105375, 2024.

\bibitem{kermany2018large}
D.~Kermany, K.~Zhang, and M.~Goldbaum, ``Large dataset of labeled optical coherence tomography (oct) and chest x-ray images,'' \emph{Mendeley Data}, vol.~3, no. 10.17632, 2018.

\end{thebibliography}

\end{document}